\documentclass[aps,groupedaddress,showpacs,floatfix]{revtex4}
\usepackage{amssymb}
\usepackage{amsmath}
\usepackage{graphicx}
\DeclareMathOperator{\Ai}{Ai}
\begin{document}

\title{On two-time distribution functions in (1+1) random directed polymers }

\author{Victor Dotsenko}

\affiliation{LPTMC, Universit\'e Paris VI, Paris, France}

\affiliation{L.D.\ Landau Institute for Theoretical Physics, Moscow, Russia}

\date{\today}

\begin{abstract}
An asymptotic analytic expression for the two-time free energy
distribution function in (1+1) random directed polymers is derived  in the limit when the
two times are close to each other.

\end{abstract}


\maketitle

\medskip

In this paper we consider  $(1+1)$  directed polymers model in terms of a scalar field $\phi(\tau)$
within an interval $0 \leq \tau \leq t$ defined by the Hamiltonian
\begin{equation}
   \label{1}
   H[\phi(\tau), V] = \int_{0}^{t} d\tau
   \Bigl\{\frac{1}{2} \bigl[\partial_\tau \phi(\tau)\bigr]^2
   + V[\phi(\tau),\tau]\Bigr\};
\end{equation}
where  $V(\phi,\tau)$ is the Gaussian distributed random potential with a zero mean, $\overline{V(\phi,\tau)}=0$,
and $\delta$-correlations, ${\overline{V(\phi,\tau)V(\phi',\tau')}} = u \delta(\tau-\tau') \delta(\phi-\phi')$.
The parameter $u$ is the strength of the disorder.
This problem, which is equivalent to the one of the KPZ-equation
\cite{KPZ} describing the time evolution of an interface
in the presence of noise, has been the focus  of intense studies during past three
decades
\cite{hh_zhang_95,burgers_74,kardar_book,hhf_85,numer1,numer2,kardar_87,bouchaud-orland,Brunet-Derrida,
Johansson,Prahofer-Spohn,Ferrari-Spohn1,KPZ-TW1a,KPZ-TW1b,KPZ-TW1c,KPZ-TW2,BA-TW1,BA-TW2,BA-TW3,
LeDoussal1,LeDoussal2,goe,LeDoussal3,Corwin,Borodin}.
At present it is well established that the fluctuations of the free energy of the model defined by the
Hamiltonian (\ref{1}) are described by the Tracy-Widom  distribution \cite{TW-GUE}.
The two-point free energy distribution function
which describes joint statistics of the free energies of the directed polymers
coming to two different endpoints has been derived in  \cite{Prolhac-Spohn,2pointPDF,Imamura-Sasamoto-Spohn}.
The generalization for the corresponding $N$-point distribution has been obtained in
\cite{N-point-1,N-point-2}.

All the above studies were devoted to the so called "one-time" situation.
The joint probability distribution function
at two different times has been studied in the paper \cite{2time-1, 2time-2}
(in terms of non-rigorous replica Bethe
ansatz approach) and in the paper \cite{2time-J} (mathematically rigorous derivation). Unfortunately,
the results obtained in these papers  are expressed in terms rather complicated mathematical objects whose analytic properties are not clear. 
Quite recently the time correlations of the KPZ interface growth has been studied both theoretically, 
in terms of the variational problem involving two independent Airy processes \cite{Ferrari-Spohn_time-corr}, 
and experimentally \cite{Takeuchi_time-corr}.

In this paper I am going to consider the two-time free energy distribution functions of the
directed polymers defined by the Hamiltonian (\ref{1}) in the limit when the two times are getting
close to each other.
For the fixed boundary conditions the partition function
of this model is
\begin{equation}
\label{2}
   Z(x; \, t) = \int_{\phi(0)=0}^{\phi(t)=x}
              {\cal D} \phi(\tau)  \;  \mbox{\Large e}^{-\beta H[\phi]}
\; = \; \exp\bigl\{-\beta F(x; t)\bigr\}
\end{equation}
where $\beta$ is the inverse temperature and $F(x,t)$ is the free energy.
In the limit $t\to\infty$ the free energy scales as
\begin{equation}
\label{3}
\beta F(x,t) = \beta f_{0} t + \beta x^{2}/2t + \lambda(t) f(x) \; ,
\end{equation}
where $f_{0}$ is a non-random free energy density,
\begin{equation}
\label{4}
\lambda(t) = \frac{1}{2}(\beta^{5} u^{2} t)^{1/3}  \propto t^{1/3}
\end{equation}
and $f(x)$ is a random quantity described by the
Tracy-Widom distribution.
As the first two trivial terms of the free energy, eq.(\ref{3}), can be easily
eliminated by simple redefinition of the partition function, they  will be
omitted in the further calculations, in other words,
\begin{equation}
\label{5}
Z(x; \, t) = \exp\bigl\{-\lambda(t) f(x)\bigr\} \; .
\end{equation}

Let us consider two directed polymer trajectories
with the fixed boundary conditions
at two different times: $\phi_{1}(0) = \phi_{1}(t) = 0$ and $\phi_{2}(0) = \phi_{2}\bigl[(1-\Delta)t\bigr] = 0$
 having the free energies $f_{1}$ and $f_{2}$ correspondingly.
By definition
\begin{equation}
\label{6}
\exp\bigl\{-\lambda(t) f_{1}\bigr\} \; = \; Z(0; \, t)
\end{equation}
and
\begin{equation}
\label{7}
\exp\bigl\{-\lambda\bigl[(1-\Delta) t\bigr] f_{2}\bigr\} \; = \; Z(0; \, t-\Delta t)
\end{equation}
where the partition function $Z(x; t)$ is defined in eq.(\ref{2}).
We are going to study the properties of the
joint probability distribution function,
$W (f_{1}, f_{2}, \Delta)$, the limit $t\to\infty$ when the parameter $0 < \Delta < 1$
remains finite.

By definition,  the probability distribution function
$W(f_{1}, f_{2}, \Delta)$ is related with the corresponding probability density
$P_{t,\Delta}(f_{1}, f_{2})$ as
\begin{equation}
\label{8}
W(f_{1}, f_{2}, \Delta) \; = \; \lim_{t\to\infty}
\int_{f_{1}}^{+\infty} \; df_{1}' \int_{f_{2}}^{+\infty} \;  df_{2}' \;
P_{t,\Delta}(f_{1}', f_{2}') \; = \;
\int_{f_{1}}^{+\infty} \; df_{1}' \int_{f_{2}}^{+\infty} \;  df_{2}' \;
{\cal P}_{\Delta}(f_{1}', f_{2}')
\end{equation}
where
\begin{equation}
\label{9}
{\cal P}_{\Delta}(f_{1}, f_{2}) \; = \; \lim_{t\to\infty} P_{t,\Delta}(f_{1}, f_{2}) \; .
\end{equation}
In terms of the above partition functions (\ref{6})-(\ref{7}) the probability distribution
function $W(f_{1}, f_{2}, \Delta)$ can be defined as follows:
\begin{equation}
\label{10}
W(f_{1}, f_{2}, \Delta) = \lim_{t\to\infty}
\sum_{N,K=0}^{\infty}
\frac{(-1)^{N+K}}{N! \; K!}
\exp\bigl(\lambda_{t} N f_{1} + \lambda_{t} K f_{2}\bigr) \;
\overline{Z^{N}(0; t) \, Z^{K}(0; t-\Delta t)}
\end{equation}
where  $\overline{(...)}$ denotes the disorder averaging  and the parameter
$\lambda(t)$ is defined in eq.(\ref{4}).
Indeed, substituting here eqs.(\ref{6}) and (\ref{7})  we get
\begin{eqnarray}
 \nonumber
W(f_{1}, f_{2}, \Delta) &=& \lim_{t\to\infty}
\sum_{N,K=0}^{\infty}
\frac{(-1)^{N+K}}{N! \; K!}
\int_{-\infty}^{+\infty} df_{1}'\int_{-\infty}^{+\infty} df_{2}'\; P_{t,\Delta}(f_{1}', f_{2}')
\exp\Bigl\{
\lambda(t) N (f_{1} - f_{1}') + \lambda[(1-\Delta)t] K (f_{2} - f_{2}') 
\Bigr\}
\\
\nonumber
\\
\nonumber
&=& \lim_{t\to\infty}
\int_{-\infty}^{+\infty} \; df_{1}' \int_{-\infty}^{+\infty} \;  df_{2}' \; P_{t,\Delta}(f_{1}', f_{2}')
\exp\Bigl\{
-\exp\bigl\{\lambda(t) (f_{1} - f_{1}')\bigr\} -
           \exp\bigl\{\lambda[(1-\Delta)t] (f_{2} - f_{2}')\bigr\} 
\Bigr\}
\\
\nonumber
\\
&=&
\int_{-\infty}^{+\infty} \; df_{1}' \int_{-\infty}^{+\infty} \;  df_{2}' \;
{\cal P}_{\Delta}(f_{1}', f_{2}') \;
\theta\bigl(f_{1}'-f_{1}\bigr) \theta\bigl(f_{2}'-f_{2}\bigr)
 \label{11}
\end{eqnarray}
which coincides with the definition, eq.(\ref{8}).

According to the definition of the partition function, eq.(\ref{2}),
\begin{equation}
 \label{12}
Z(0; \, t) \; = \;
\int_{-\infty}^{+\infty} dx \;
Z(x; \, t-\Delta t)\, Z^{*}(x; \, \Delta t)
\end{equation}
where $Z^{*}(x; \, \Delta t)$ is the partition function of the directed polymer system
in which time goes backwards, from $t$ to $(t - \Delta t)$. Substituting eq.(\ref{12})
into eq.(\ref{10}) we get
\begin{eqnarray}
\nonumber
W(f_{1}, f_{2}, \Delta) &=& \lim_{t\to\infty}
\sum_{N,K=0}^{\infty}
\frac{(-1)^{N+K}}{N!\, K!}
\exp\Bigl[\lambda(t) N f_{1} + \lambda(t-\Delta t) K f_{2}\Bigr] \times
\\
\nonumber
\\
&\times&
\int_{-\infty}^{+\infty} dx_{1} \, ... \, dx_{N}
\overline{\Bigl[\prod_{a=1}^{N} Z(x_{a}; \, t-\Delta t)\Bigr] Z^{K}(0; \, t-\Delta t)
          \Bigl[\prod_{a=1}^{N} Z^{*}(x_{a}; \, \Delta t)\Bigr]}
\label{13}
\end{eqnarray}
or
\begin{eqnarray}
\nonumber
W(f_{1}, f_{2}, \Delta) &=& \lim_{t\to\infty}
\sum_{N,K=0}^{\infty}
\frac{(-1)^{N+K}}{N!\, K!} \,
\exp\Bigl[\lambda(t) N f_{1} + \lambda(t-\Delta t) K f_{2}\Bigr] \times
\\
\nonumber
\\
&\times&
\int_{-\infty}^{+\infty} dx_{1}...dx_{N}
\Psi\bigl(x_{1},...,x_{N}, \underbrace{0, ..., 0}_{K} ; \; (t - \Delta t) \bigr) \;
\Psi^{*}\bigl(x_{1},...,x_{N} ; \; \Delta t\bigr)
\label{14}
\end{eqnarray}
where
\begin{equation}
\label{15}
\Psi(x_{1}, ..., x_{N} ; \; t) \; \equiv \;
\overline{Z(x_{1}; \, t) \, Z(x_{2} \, t) \, ... \, Z(x_{N} \, t)}
\end{equation}
One can easily show that $\Psi({\bf x}; \, t)$ is the wave function of $N$-particle quantum
boson system with attractive $\delta$-interaction:
\begin{equation}
   \label{16}
\beta \, \partial_t \Psi({\bf x}; t) =
\frac{1}{2}\sum_{a=1}^{N}\partial_{x_a}^2 \Psi({\bf x}; t)
+\frac{1}{2}\, \beta^{3} u \sum_{a\not=b}^{N} \delta(x_a-x_b) \; \Psi({\bf x}; t)
\end{equation}
with the initial condition $\Psi({\bf x}; \, 0) = \Pi_{a=1}^{N} \delta(x_a)$.
The time dependent wave function  $\Psi({\bf x}; \, t)$ of this quantum problem can be represented
in terms of the linear combination of the Bethe ansatz eigenfunctions  of eq.(\ref{16})
which are well known since long time ago
(for details see e.g.\cite{Lieb-Liniger,McGuire,Yang,Calabrese,rev-TW})


Using this representation the general (rather cumbersome) result for $W(f_{1}, f_{2}, \Delta)$,
eq.(\ref{14})  (for an arbitrary $\Delta$) has been  derived in \cite{2time-1}
(see eqs(73), (76) and (79)-(82)).
In this paper we would like to study the limit $\Delta \to 0$ which is somewhat tricky
as it is considered {\it after} the limit $t \to \infty$ is already taken.

It is clear that at $\Delta = 0$ the two free energies $f_{1}$ and $f_{2}$ must coincide, so that
$ {\cal P}_{\Delta=0}(f_{1}; f_{2}) \; = \; \delta( f_{1}- f_{2}) \; P_{GUE}(f_{1})$, and
$W(f_{1}, f_{2}, \Delta=0) \; = \; F_{2} (-\mbox{max}\{f_{1}, f_{2}\})$ where $F_{2} (-f)$ is the
GUE Tracy-Widom distribution. However, when $\Delta \ll 1$ but not zero, the situation becomes much less
trivial. In this case the distribution density $ P_{\Delta}(f_{1}; f_{2}) $ have a shape of a narrow
"$\delta$-like" peak with a width $\sim \Delta^{1/3}$ around $f_{1}-f_{2}$.
Then, if $|f_{1}- f_{2}| \gg \Delta^{1/3}$,  the "$\delta$-like" distribution
$ P_{\Delta}(f_{1}; f_{2}) $ can be treated as the $\delta$-function,
so that one would get $W(f_{1}, f_{2}, \Delta) \simeq F_{2} (-\mbox{max}\{f_{1}, f_{2}\})$ again.
However, if $|f_{1}- f_{2}| \sim \Delta^{1/3}$
the situation becomes much more complicated.

Let us introduce a new scaling variable
\begin{equation}
\label{17}
\xi \; = \; \frac{f_{1}-f_{2}}{\Delta^{1/3} \, 2^{2/3}}
\end{equation}
where the factor $2^{2/3}$ is introduced just to simplify the final formula.
Substituting $f_{2} =  2^{2/3} \,f $ and
$f_{1} =  2^{2/3} \, f \; + \; \Delta^{1/3} \, 2^{2/3} \, \xi $ into
$W(f_{1}, f_{2}, \Delta)$ and taking the limit $\Delta \to 0$ (so that $\xi$ remains finite)
we define a new probability distribution function
\begin{equation}
\label{18}
W(f, \xi) \; \equiv \;
\lim_{\Delta\to 0} W\Bigl(2^{2/3} f \, + \,  2^{2/3} \Delta^{1/3} \, \xi, \; \; 2^{2/3} f ; \; \Delta\Bigr)
\end{equation}
which describe the "detailed shape" of the distribution  $W(f_{1}, f_{2}, \Delta)$
in the limit $\Delta \to 0$ when the two free energies $f_{1}$ and $f_{2}$ are close to each other
$(f_{1}- f_{2}) \; \sim \; \Delta^{1/3}$.

It can be shown (see Appendix) that the result for $W(f, \xi)$ can be represented in
terms of two Fredholm determinants in the form of the following rather compact analytic formula:
\begin{equation}
\label{19}
W(f, \xi) \; = \;
\frac{1}{2\pi i} \int_{|z|=1} \frac{dz}{z} \det\Bigl[\hat{1} \; - \; (1-z) \hat{K}_{-f}\Bigr] \;
                                           \det\Bigl[\hat{1} \; + \; z^{-1} \hat{K}_{-\xi}\Bigr]
\end{equation}
where the integration over $z$ goes in the complex plane around zero and $\hat{K}_{-f}$ is the
integral operator with the usual Airy kernel $K(u-f; u'-f)$   (with $(u, \, u') \; \geq 0$ ):
\begin{equation}
\label{20}
K(u-f; u'-f) \; = \; \int_{0}^{+\infty} dy \; \Ai(y + u - f) \; \Ai(y + u' - f)
\end{equation}
Following the standard transformations of the Fredholm determinant with the Airy kernel
\cite{TW-GUE} the distribution function, eq.(\ref{19}),
can equivalently be represented as
\begin{equation}
\label{21}
W(f, \xi) \; = \;
\frac{1}{2\pi i} \int_{|z|=1} \frac{dz}{z}
\exp\Biggl\{
-\int_{-f}^{\infty} dt \, (t+f) \; q_{1}^{2}(t) \; - \;
\int_{-\xi}^{\infty} dt \, (t+\xi) \; q_{2}^{2}(t)
\Biggr\}
\end{equation}
where $q_{1}(t)$ and $q_{2}(t)$ are the solutions of the Panlev\'e II
equation
\begin{equation}
 \label{22}
q''(t) = t q(t) + 2 q^{3}(t)
\end{equation}
with the boundary conditions, $q_{1}(t\to +\infty) \sim (1-z)^{1/2} \, \Ai(t)$
and $q_{2}(t\to +\infty) \sim (-z)^{-1/2} \, \Ai(t)$ correspondingly.
Eqs.(\ref{19})-(\ref{22})  constitute  the main result of the present paper.
The analytic properties of the function $W(f, \xi)$ given by the above formulas
are still to be investigated.

\acknowledgments

I am grateful to Alexei Borodin,  Kurt Johansson, Patrick Ferrari, Jinho Baik, Kazumasa Takeuchi, Kostya Khanin
Herbert Spohn and Craig Tracy for numerous illuminating discussions.
An essential part of this work was done during the workshop "New approaches to non-equilibrium
and random systems: KPZ integrability, universality, applications and experiments"
(Jan 11 - Mar 11, 2016)
at Kavli Institute of Theoretical Physics, University of California, Santa Barbara.
This research was supported in part by the National Science Foundation under Grant No. NSF PHY11-25915.

\vspace{15mm}

\begin{center}

\appendix{\Large Appendix }

\end{center}

\newcounter{A}
\setcounter{equation}{0}
\renewcommand{\theequation}{A.\arabic{equation}}

\vspace{5mm}

In the notations of the present paper the general result of the paper \cite{2time-1},
(eqs(73), (76) and (79)-(82)) for the probability distribution function, eq.(\ref{14}),
is given by
\begin{eqnarray}
 \label{A1}
W(f_{1},f_{2},\Delta) &=&
\sum_{M_{1}=0}^{\infty} \frac{(-1)^{M_{1}}}{(M_{1}!)^{2}}
\sum_{M_{2}=0}^{\infty} \frac{(-1)^{M_{2}}}{M_{2}!}
\prod_{\alpha=1}^{M_{1}+M_{2}}
\Biggl[
\int_{0}^{\infty} du_{\alpha}
\Biggr]
\;
\prod_{\beta=1}^{M_{1}}
\Biggl[
\int_{0}^{\infty} dv_{\beta}
\Biggr]
\times
\\
\nonumber
\\
\nonumber
&\times&
\sum_{{\cal P}\in S_{M_{1}+M_{2}}} (-1)^{\bigl[{\cal P}\bigr]}
\sum_{\tilde{{\cal P}}\in S_{M_{1}}} (-1)^{\bigl[\tilde{{\cal P}}\bigr]}
\prod_{\beta=1}^{M_{1}}
\Bigl[
G\bigl(u_{\beta}, v_{\beta}; \; u_{{\cal P}_{\beta}},  v_{\tilde{{\cal P}}_{\beta}}\bigr)
\Bigr]
\times
\\
\nonumber
\\
\nonumber
&\times&
\prod_{\gamma=1}^{M_{2}}
\Bigl[
2^{1/3}K\bigl[2^{1/3} u_{M_{1} + \gamma} - \tilde{f}_{2} ; \;
              2^{1/3} u_{{\cal P}_{M_{1} + \gamma}} - \tilde{f}_{2} \bigr]
\Bigr]
\end{eqnarray}
where ${\cal P}$ and $\tilde{{\cal P}}$ denote the permutations of $(M_{1}+M_{2})$ variables
$\{u_{\alpha}\}$ and $M_{1}$ variables $\{v_{\beta}\}$ correspondingly,
$K(u; u')  =  \int_{0}^{+\infty} dy \; \Ai(y + u) \; \Ai(y + u')$
is the Airy kernel and
\begin{equation}
 \label{A2}
G\bigl( u, u' ; \; v, v'\bigr) \; = \;
\sum_{i=1}^{4} G_{i}\bigl( u, u' ; \; v, v'\bigr)
\end{equation}
with
\begin{eqnarray}
 \label{A3}
G_{1}\bigl( u, u' ; \; v, v'\bigr) &=&
2^{2/3}\bigl( 1 - \Delta\bigr)^{2/3}
\Ai\Biggl[2^{1/3}\bigl( 1 - \Delta\bigr)^{1/3} (u + v') - \tilde{f}_{1}\Biggr] \,
\Ai\Biggl[2^{1/3}\bigl( 1 - \Delta\bigr)^{1/3} (u' + v) - \tilde{f}_{1}\Biggr] \,
\\
\nonumber
\\
\nonumber
\\
\nonumber
G_{2}\bigl( u, u' ; \; v, v'\bigr)  &=&
 -2^{2/3}\Bigl(\frac{1-\Delta}{\Delta}\Bigr)^{1/3}
K\Bigl[2^{1/3} u - \tilde{f}_{2}; \; 2^{1/3} u' - \tilde{f}_{2} \Bigr]
\times
\\
\nonumber
\\
&\times&
K\Biggl[
2^{1/3}\Bigl(\frac{1-\Delta}{\Delta}\Bigr)^{1/3} v -
     \frac{\tilde{f}_{1} - (1-\Delta)^{1/3} \tilde{f}_{2}}{\Delta^{1/3}} \, ; \;
2^{1/3}\Bigl(\frac{1-\Delta}{\Delta}\Bigr)^{1/3} v' -
     \frac{\tilde{f}_{1} - (1-\Delta)^{1/3} \tilde{f}_{2}}{\Delta^{1/3}}
\Biggr]
\label{A4}
\end{eqnarray}
\begin{eqnarray}
\nonumber
G_{3}\bigl( u, u' ; \; v, v'\bigr)  &=&
 -2 \Bigl(\frac{1-\Delta}{\Delta}\Bigr)^{1/3}
\int_{-\infty}^{+\infty} \frac{dp}{2\pi}
\int_{0}^{+\infty} dy \;
\times
\\
\nonumber
\\
\nonumber
&\times&
\Ai\Biggl[
-\Bigl(\frac{1-\Delta}{\Delta}\Bigr)^{1/3} y
+ \Bigl(\frac{1-\Delta}{\Delta}\Bigr)^{-2/3} p^{2}
- \frac{f_{1} - (1-\Delta)^{1/3} f_{2}}{\Delta^{1/3}}
+ \Bigl(\frac{1-\Delta}{\Delta}\Bigr)^{1/3}(v + v')
\Biggr]
\times
\\
\nonumber
\\
&\times&
\Ai\bigl(y + p^{2} - f_{2} + u + u'\bigr)
\exp\{ip \, (u-u'-v+v')\}
\label{A5}
\end{eqnarray}
\begin{eqnarray}
\nonumber
G_{4}\bigl( u, u' ; \; v, v'\bigr)  &=&
 -4 \Bigl(\frac{1-\Delta}{\Delta}\Bigr)^{1/3}
\int\int_{-\infty}^{+\infty} \frac{dp \, dq}{(2\pi)^{2}}
\int_{0}^{+\infty} dy \;
\times
\\
\nonumber
\\
\nonumber
&\times&
\Ai\Biggl[
-\Bigl(\frac{1-\Delta}{\Delta}\Bigr)^{1/3} y
+ \Bigl(\frac{1-\Delta}{\Delta}\Bigr)^{-2/3} p^{2}
- \frac{f_{1} - (1-\Delta)^{1/3} f_{2}}{\Delta^{1/3}}
+ \Bigl(\frac{1-\Delta}{\Delta}\Bigr)^{1/3}(v + v')
\Biggr]
\times
\\
\nonumber
\\
&\times&
\Ai\bigl(y + q^{2} - f_{2} + u + u'\bigr)
\frac{\sin\bigl[(q-p) \, y\bigr]}{(q-p)}
\exp\{iq(u-u') - ip(v-v')\}
\label{A6}
\end{eqnarray}
where $\tilde{f}_{1,2} \equiv  2^{-2/3} \, f_{1,2} $

Introducing the new scaling parameter $\xi$, eq.(\ref{17}),
and taking the limit $\Delta \to 0$ (such that the parameter $\xi$ remains finite)
we get:
\begin{eqnarray}
 \label{A8}
G_{1}\bigl( u, u' ; \; v, v'\bigr)\Big|_{\Delta\to 0} &\simeq&
2^{2/3}
\Ai\Bigl[2^{1/3}(u + v') - \tilde{f}_{1}\Bigr] \,
\Ai\Bigl[2^{1/3}(u' + v) - \tilde{f}_{1}\Bigr] \,
\\
\nonumber
\\
\nonumber
\\
G_{2}\bigl( u, u' ; \; v, v'\bigr)\Big|_{\Delta\to 0} &\simeq&
 -\frac{2^{2/3}}{\Delta^{1/3}}
K\Bigl[2^{1/3} u - \tilde{f}_{2}; \; 2^{1/3} u' - \tilde{f}_{2} \Bigr]
\;
K\Biggl[
\frac{2^{1/3} v}{\Delta^{1/3}} - \xi \, ; \;
\frac{2^{1/3} v'}{\Delta^{1/3}} - \xi \,
\Biggr]
\label{A9}
\end{eqnarray}
Redefining $y \to \Delta^{1/3} y$ in eq.(\ref{A5}) and taking the limit $\Delta \to 0$
we obtain
\begin{eqnarray}
\nonumber
G_{3}\bigl( u, u' ; \; v, v'\bigr)\Big|_{\Delta\to 0} &\simeq&
 -2
\int_{-\infty}^{+\infty} \frac{dp}{2\pi}
\int_{0}^{+\infty} dy \;
\times
\\
\nonumber
\\
&\times&
\Ai\Biggl(
- y - \tilde{\xi} + \frac{v + v'}{\Delta^{1/3}}
\Biggr)
\;
\Ai\bigl(p^{2} - f_{2} + u + u'\bigr)
\exp\{ip \, (u-u'-v+v')\}
\label{A10}
\end{eqnarray}
Finally, redefining $y \to \Delta^{1/3} y$, $p \to \Delta^{-1/3} p$
in eq.(\ref{A6}) and taking the limit $\Delta \to 0$
we find
\begin{eqnarray}
\nonumber
G_{4}\bigl( u, u' ; \; v, v'\bigr)\Big|_{\Delta\to 0} &\simeq&
 -4
\int\int_{-\infty}^{+\infty} \frac{dp \, dq}{(2\pi)^{2}}
\int_{0}^{+\infty} dy \;
\frac{\sin\bigl(p \, y\bigr)}{p}
\times
\\
\nonumber
\\
&\times&
\Ai\Biggl(
- y +  p^{2} - \tilde{\xi} + \frac{v + v'}{\Delta^{1/3}}
\Biggr)
\;
\Ai\bigl(q^{2} - f_{2} + u + u'\bigr)
\exp\{iq(u-u') - ip(v-v')\}
\label{A11}
\end{eqnarray}

Redefining $v_{\beta} \to \Delta^{1/3} \, 2^{-1/3} \, v_{\beta}$ and
$u_{\alpha} \to  2^{-1/3} \, u_{\alpha}$ in eq.(\ref{A1}) we find that
the kernel
$G\bigl( u, u' ; \; v, v'\bigr) \; = \; \sum_{i=1}^{4} G_{i}\bigl( u, u' ; \; v, v'\bigr)$
gets the multiplicative factor $\Delta^{1/3}$.
According to eqs.(\ref{A8}), (\ref{A10}) and (\ref{A11}), we find that
\begin{equation}
\label{A12}
\Delta^{1/3} \, G_{1,3,4}\bigl( u, u' ; \; v, v'\bigr)\big|_{\Delta\to 0} \; \to \; 0
\end{equation}
According to eq.(\ref{A9}), the only contribution which remains finite in the limit
$\Delta \, \to \, 0$ is:
\begin{equation}
 \label{A13}
\Delta^{1/3} \, G_{2}\bigl( u, u' ; \; v, v'\bigr) \;  \to \;
 - K\bigl(u - \tilde{f}_{2}; \; u' - \tilde{f}_{2} \bigr)
\;
K\bigl(v - \xi \, ; \; v' - \xi \, \bigr)
\end{equation}
Thus, substituting $\tilde{f}_{2} = f$,
$\; \tilde{f}_{1} \; = \; f \, + \, \Delta^{1/3} \xi$,
$\; v_{\beta} \; \to \Delta^{1/3} \, 2^{-1/3} \, v_{\beta}$ and
$u_{\alpha} \to  2^{-1/3} \, u_{\alpha}$ in eq.(\ref{A1}), in the limit $\Delta \, \to \, 0$
we get

\begin{eqnarray}
 \nonumber
 W(f, \xi) & \equiv &
\lim_{\Delta\to 0} W\Bigl(2^{2/3} f \, + \,  2^{2/3} \Delta^{1/3} \, \xi, \; \; 2^{2/3} f ; \; \Delta\Bigr)
\\
\nonumber
\\
\nonumber
&=&
\sum_{M_{1}=0}^{\infty} \frac{(-1)^{M_{1}}}{(M_{1}!)^{2}}
\sum_{M_{2}=0}^{\infty} \frac{(-1)^{M_{2}}}{M_{2}!}
\prod_{\alpha=1}^{M_{1}+M_{2}}
\Biggl[
\int_{0}^{\infty} du_{\alpha}
\Biggr]
\;
\prod_{\beta=1}^{M_{1}}
\Biggl[
\int_{0}^{\infty} dv_{\beta}
\Biggr] \;
\sum_{{\cal P}\in S_{M_{1}+M_{2}}} (-1)^{\bigl[{\cal P}\bigr]}
\sum_{\tilde{{\cal P}}\in S_{M_{1}}} (-1)^{\bigl[\tilde{{\cal P}}\bigr]}
\times
\\
\nonumber
\\
\nonumber
&\times&
\prod_{\beta=1}^{M_{1}}
\Bigl[
- K\bigl(u_{\beta} - f; \; u_{{\cal P}_{\beta}} - f \bigr)
\;
K\bigl(v_{\beta} - \xi \, ; \; v_{\tilde{{\cal P}}_{\beta}} - \xi \, \bigr)
\Bigr]
\; 
\prod_{\gamma=1}^{M_{2}}
\Bigl[
K\bigl( u_{M_{1} + \gamma} - f ; \; u_{{\cal P}_{M_{1} + \gamma}} - f \bigr)
\Bigr]
\label{A14}
\\
\nonumber
\\
\nonumber
\\
\nonumber
&=&
\sum_{M_{1}=0}^{\infty} \frac{1}{(M_{1}!)^{2}}
\sum_{M_{2}=0}^{\infty} \frac{(-1)^{M_{2}}}{M_{2}!}
\prod_{\alpha=1}^{M_{1}+M_{2}}
\Biggl[
\int_{0}^{\infty} du_{\alpha}
\Biggr]
\;
\prod_{\beta=1}^{M_{1}}
\Biggl[
\int_{0}^{\infty} dv_{\beta}
\Biggr]
\times
\\
\nonumber
\\
&\times&
\sum_{{\cal P}\in S_{M_{1}+M_{2}}} (-1)^{\bigl[{\cal P}\bigr]}
\prod_{\alpha=1}^{M_{1}+M_{2}}
\Bigl[
K\bigl(u_{\alpha} - f; \; u_{{\cal P}_{\alpha}} - f \bigr)
\Bigr]
\sum_{\tilde{{\cal P}}\in S_{M_{1}}} (-1)^{\bigl[\tilde{{\cal P}}\bigr]}
\prod_{\beta=1}^{M_{1}}
\Bigl[
K\bigl(v_{\beta} - \xi \, ; \; v_{\tilde{{\cal P}}_{\beta}} - \xi \, \bigr)
\Bigr]
\label{A15}
\end{eqnarray}
or
\begin{eqnarray}
 \nonumber
W(f, \xi) & = &
\sum_{M=0}^{\infty} \frac{(-1)^{M}}{M!}
\prod_{\alpha=1}^{M}
\Biggl[
\int_{0}^{\infty} du_{\alpha}
\Biggr]
\det
\Bigl(
K\bigl[ u_{\alpha} - f ; \;u_{\alpha'} - f \bigr]
\Bigr)\Big|_{(\alpha,\alpha')=1,...,M}
\times
\\
\nonumber
\\
&\times&
\sum_{M_{1}=0}^{M} \frac{(-1)^{M_{1}}M!}{(M_{1}!)^{2} (M-M_{1})!}
\prod_{\beta=1}^{M_{1}}
\Biggl[
\int_{0}^{\infty} dv_{\beta}
\Biggr]
\det
\Bigl(
K\bigl[ v_{\beta} - \xi ; \;v_{\beta'} - \xi \bigr]
\Bigr)\Big|_{(\beta,\beta')=1,...,M_{1}}
\label{A16}
\end{eqnarray}
It can be easily shown (see below) that the above formula can be represented in the compact
form in terms of the product of two Fredholm determinants
\footnote{I am grateful to Alexei Borodin for his indication that eq.(\ref{A16}) can be
represented in the form of eq.(\ref{A17})}
\begin{equation}
\label{A17}
W(f, \xi) \; = \;
\frac{1}{2\pi i} \int_{|z|=1} \frac{dz}{z} \det\Bigl[\hat{1} \; - \; (1-z) \hat{K}_{-f}\Bigr] \;
                                           \det\Bigl[\hat{1} \; + \; z^{-1} \hat{K}_{-\xi}\Bigr]
\end{equation}
where the integration over $z$ goes in the complex plane around zero and $\hat{K}_{-f}$ is the
integral operator with the usual Airy kernel $K(u-f; u'-f)$  (with $(u, \, u') \; \geq 0$).
Indeed, expanding the Fredholm determinants in eq.(\ref{A17}), one gets
\begin{equation}
\label{A18}
W(f, \xi) \; = \;
\frac{1}{2\pi i} \int_{|z|=1} \frac{dz}{z}
\sum_{M=0}^{\infty} \frac{(-1)^{M}}{M!} \, (1-z)^{M} D_{M}(-f) \; 
\sum_{M_{1}=0}^{\infty} \frac{1}{M_{1}!} \, z^{-M_{1}} D_{M_{1}}(-\xi)
\end{equation}
where
\begin{equation}
\label{A19}
D_{M}(s) \; \equiv \; 
\prod_{\alpha=1}^{M}
\Biggl[
\int_{0}^{\infty} du_{\alpha}
\Biggr]
\det
\Bigl(
K\bigl[ u_{\alpha} + s ; \;u_{\alpha'} + s \bigr]
\Bigr)\Big|_{(\alpha,\alpha')=1,...,M}
\end{equation}
Simple algebra yields:
\begin{eqnarray}
 \nonumber
W(f, \xi) &=& 
\sum_{M=0}^{\infty} \frac{(-1)^{M}}{M!} D_{M}(-f)
\sum_{M_{1}=0}^{\infty} \frac{1}{M_{1}!} D_{M_{1}}(-\xi)
\sum_{L=0}^{M} \frac{(-1)^{L} M!}{L! (M-L)!} \; 
\frac{1}{2\pi i} \int_{|z|=1} \frac{dz}{z} \; z^{L - M_{1}}
\\
\nonumber
\\
&=&
\sum_{M=0}^{\infty} \frac{(-1)^{M}}{M!} D_{M}(-f)
\sum_{M_{1}=0}^{\infty} \frac{1}{M_{1}!} D_{M_{1}}(-\xi)
\sum_{L=0}^{M} \frac{(-1)^{L} M!}{L! (M-L)!} \; \delta_{L, M_{1}}
\label{A20}
\end{eqnarray}
which coinsides with eq.(\ref{A16}).

\end{document}